\documentstyle[12pt]{article}
\begin{document} \thispagestyle{empty}

\noindent\hfill  OHSTPY-HEP-TH-94-19\\
\begin{center}\begin{Large}\begin{bf}
Topology and Confinement In Light-Front QCD\footnote{invited talk at the
workshop on
 "Theory of Hadrons ond Light-Front QCD" at Polona Zgorzelisko, Poland August
15-25 1994}\\
\end{bf}\end{Large}\end{center}
\vspace{.75cm}\begin{center}
Stephen Pinsky\\[10pt]
     \end{center}
      \vspace{0.1cm}
      \begin{center}
      \begin{it}
Department of Physics\\
The Ohio State University\\
174 West 18th Avenue\\
Columbus, Ohio  43210\\
       \end{it}
         \end{center}
\vspace{1cm} \baselineskip=35pt \begin{abstract} \noindent
In 1+1 dimensional compact QCD the zero modes of $A^+$ give the theory a
non-trivial topological  structure. We examine the effects of these
topological structures on the confining infrared structure of the
theory. We show that the ground state wavefunction of the topological
excitation smears the infrared behavior sufficiently to eliminate
confinement for some matter currents. We review the work of Franke et.
al. \cite{Franke} which shows that the zero modes of
$A^+$ in $QCD_{3+1}$ give rise to instantons. The relation of zero modes of
$A^+$,
instantons and confinement in $QCD_{3+1}$ is discussed.

\end{abstract}
\newpage

{\bf 1. Introduction}
\vspace{.1in}
\baselineskip=18pt

\noindent It is almost universally accepted that Quantum Chromodynamics (QCD)
is the correct theory of the strong interactions. There are
numerous reasons for
this wide-spread acceptance of QCD.  In particular the
comparison between perturbative QCD and experiments is generally
very convincing.
There are few if any tests however that probe the strong coupling
regime of the theory.
It is crucial that a successful non-perturbative solution expose
the three important long range properties of QCD:  confinement,
spontaneous chiral symmetry breaking, and the topological structure.
In the
limit of a vanishing mass matrix, the chiral $U(N_f)\otimes U(N_f)$ symmetry is
explicit in the QCD Lagrangian. To avoid parity doublets in the spectrum, this
chiral symmetry must be spontaneously broken to $SU(N_f)$ leaving
$N_f^2-1$ Goldstone particles to be identified with the pions.
Any successful calculation must produce this symmetry breaking.
QCD appears to have an
extra $U_A(1)$ symmetry and if QCD actually had this additional symmetry it
would
lead to one more Goldstone particle the $\eta$ and force
$ m_\eta \le \sqrt{3} m_\pi$
\cite{wein}.  To avoid this $U(1)_A$ problem the solution of QCD must have
the third
long range property, topological structure, that breaks the $U_A(1)$ symmetry
\cite{thooft}.

The calculation of the pseudo-scalar meson  spectrum from
first principles is a true test of QCD becauses it requires all three of
these  long range properties.
The light-meson calculation should be considered the ``hydrogen-atom
calculation'' for QCD. It is not
sufficient to simply put the QCD action into a large computer and then look at
the answer. We must have a calculation that provides some overall
physical insight into the way nonperturbative, nonlinear
processes work in this theory \cite{BPV}. It is the objective of the
light-front (LF)
QCD program to provide such a formalism. We believe that in LF formalism
it is possible to isolate the degrees of
freedom that are responsible for the long-range properties of QCD, and
obtain a deeper understanding of QCD.
All of the long range physics required by QCD have been discussed in great
detail in this workshop. There is every reason to expect a very strong
interplay between these long range properties. If topology strongly affects the
mass of the $\eta$ then it is logical that it plays a role in confinement as
well. We will illustrate here how topology can play a role in confinement in LF
QCD.

Perry and Wilson (to appear in
these proceedings) argue that the confinement mechanism of
$QCD_{1+1}$ can be promoted to a three dimensional confinement mechanism
through
non-perturbative renormalization in a LF quantized formulation of QCD.
Their point is that
$QCD_{3+1}$  already has a confining interaction term in the LF
Hamiltonian, the instantaneous four Fermion interaction, which is the
confining interaction in $QCD_{1+1}$. The issue, as they see it, is to
show that the second order quark glue interaction does not cancel the
instanteous interaction as it does in perturbation theory. I will point
out that the instantaneous interaction of the off diagonal currents is
modified by the topological properties of the theory, and confinement is
destroyed for these currents in $QCD_{1+1}$ on a cylinder. In
$QCD_{1+1}$ on a cylinder the topological structure is carried by the zero
mode (ZM)
of $A^+$
\cite{kalloniatis}. I argue that a similar effect is expected in
$QCD_{3+1}$  where the topology is just that associated with the instanton
and is not
an artifact of the space in which  the theory is formulated.

\vskip.3in
{\bf 2. Topology and Confinement in Two Dimensions}
\vskip.1in

\noindent In a two dimensional $SU(N)$ gauge field theory in flat space the
gauge
field  provides a confining intersection for the matter degrees of freedom.
This
is evident from the form of the light-front Hamiltonian%
\begin{equation}%
P^- = -g^2\int dx Tr(J^+ {1\over(\partial_-)^2}J^+)
\end{equation}%
The operator $1/(\partial_-)^2$ becomes $1/(k^+)^2$ in momentum space
and provides the linearly confining potential.

The theory develops a topological gluon degree of freedom when the system
is put in a
light-front spatial box with period boundary conditions. This degree of
freedom is
the ZM of $A^+$ \cite{kalloniatis}. The existence of the topological degree of
freedom hinges on the fact that the space is in compact $x^-$. When one
attempts to completely fix the gauge one finds that the allowed gauge
transformation
must be periodic, up to an element of $Z_N$, to maintain the period
boundary conditions on the fields. Thus the field can not be brought to
$A^+=0$. The best that one can do is  $\partial_- A^+ =0$.  An
additional global gauge rotation can be made on $A^+$  to bring it to diagonal
form.  Thus for $SU(2)$, we have in the end a single color comonet as a degree
of freedom

\begin{equation}%
{q(x^+) \over2L} \delta_{a,3} ={1\over2L}\int_{0}^{2L} {dx^-
A^+_a(x^+,x^-)}.
\end{equation}%

\noindent This seemingly harmless rotation to a diagonal basis has significant
consequences. With this gauge fixing only the off diagonal part of Gauss's
Law can
be solve strongly. The diagonal part must be imposed as a condition on the
states. In $QCD_{3+1}$  this is not always desirable.
In $QCD_{1+1}$, $q(x^+)$ is
the only gluon degree of freedom and the exponential of q is just the
Wilson loop
around the cylindrical space that the theory lives in.
This degree of freedom enters the Hamiltonian in several places. There is a
kinematic term $\pi^2_q/2$ and the potential $1/(k^+)^2$ between the
off-diagonal matter currents becomes
\begin{equation}%
{1\over{(k^+  \pm g (q/2L))^2}}.
\end{equation}%

\noindent The interaction potential between the diagonal currents is
unaffected.
The states of the theory are the normal Fock states of the matter fields
tensored with the states associated with the $\pi$ and $q$ operators. The
states associated with
$\pi_q$ and $q$  are described in the Schordinger picture by wavefunctions and
energy levels (see Kalloniatis in these proceedings)

\begin{equation}%
\phi_n = C_n \sin((n+1)gq) \;\;\;  E_n= ((n-1)^2-1)g^2 \pi^2 2L/8.
\end{equation}%

\noindent Thus the complete states can be written
\begin{equation}%
|n;N_i> = C_n\sin((n+1)gq)|N_i>
\end{equation}%

\noindent where $N_i$ denotes the Fock space states of the matter field
which are
operated on by the  matter currents. Very similar results are found inn equal
time quantization\cite{cyl}. Matrix elements are now of the form
\begin{equation}%
\int_{0}^{2pi/g} dq <n;N_i|"operator"|m;N_j>
\end{equation}%

\noindent where the $q$ integral runs over one fundamental modular domain
\cite{vBa92}.

The energy level splitting of the $q$ states are of order $2L$, and  in the
large $L$ limit we expect only the ground state to contribute. The ground state
wavefunction is not trivial and the matter fields will be affected by it.
Consider the matrix element of the  off diagonal currents in the
Hamiltonian between states with $N_i$ and $N_j$ matter excitations.  In
momentum space the matrix element takes the form;
\begin{equation}%
\sum_{n} <N_i|J(k_n) J(-k_n)|Nj> \{\int_0 ^{2\pi} {\sin(x)^2 dx\over{(2 \pi n +
x)^2}} + \int_0 ^{2\pi} {\sin(x)^2 dx\over{(2 \pi n -
x)^2}} \}
\end{equation}%

\noindent where $x=gq$ and $k_n = 2 \pi n/2L$.
Doing the integral we find for small momentum
\begin{equation}%
 \sum_{n}{<N_i|J(k_n) J(-k_n)|Nj>\over{.67 +O(n^2)}}
\end{equation}%

The potential now has a form similar to that of the exchange of a massive
particle.
The topological ground state has smeared out the infrared behavior and we
do not have
a confining interaction for the off diagonal currents. The diagonal currents
are
 unaffected by the topological degree of freedom.

\vskip.1in
{\bf The 3. $A^+$ ZM in 3+1 Dimensions}
\vskip.1in

\noindent Many years ago by Franke, Novozhilov and Prokhvatilov \cite{Franke}
showed  that there must be a ZM of $A^+$ in LC $QCD_{3+1}$  for the same reason
that there must be a ZM of $A^+$ in  LC $QCD_{1+1}$. Again the ZM of
$A^+$ is a dynamical variable and again it carries the topological
properties  of the theory. They only imposed periodic boundary conditions
in the $x^-$ direction and their gauge condition was $\partial_- A^+
=0$. Now the topological properties are those of the well known
instanton.

The instanton winding number can be written in terms of the
topological  current  $K^\mu$ (which is a function of $A^\mu$) as;
\begin{equation}%
\nu = {g^2\over 8\pi^2} \int (K^+(T) - K^+(0)) dx^- d^2x^\perp
\end{equation}%

\noindent where $T$ is a large LC time. The winding number shows the
relation between two pure gauge configurations separated by a large LC time
$T$. These two pure gauge configuration must be related by a gauge
transformation.  Thus we are looking for transformations that take one from one
pure ZM configuration to another pure ZM configuration. All such gauge
transforms give an integer result for the winding number. Those transformations
that give non-zero results are essentially the instantons of the theory. An
example of such a gauge transformation is
\begin{equation}%
G(x^+,x^-,x^\perp)=Exp({iN\pi\over{L}}\;\vec\sigma\cdot\vec n\; x^-)
\end{equation}%

\noindent where $\vec n $ is a space dependent unit vector and %
\begin{equation}%
\vec\sigma\cdot\vec n={{-2ax_2\sigma_1
+2ax_1\sigma_2+(a^2-x_1^2-x_2^2)}\over{(a^2+x_1^2+x_2^2)}}
\end{equation}%

\noindent We see the usual instanton structure that mix color and space
indices and
$"a"$ can be interpreted as the instanton size.

The ZM of $A^+$ enters the Hamiltonian in $QCD_{3+1}$ through structures
similar to those we saw in $QCD_{1+1}$.  Again the the ZM of
$A^+$ is related to the topology. In 3+1 the topology is not related to the
compact spatial manifold but is related to the fact that there are non-trivial
gauge transforms that leave us in the gauge
$\partial_- A^+_a =0$. These transformation do not preserve the diagonal
gauge for
$A^+$.  Within a given winding number sector one can use the diagonal gauge but
if one wants to consider processes that change winding number one must relax
this condition.

We have seen that the two dimensional confinement mechanism seen in
$QCD_{1+1}$ can be profoundly changed by the topological structure of the
theory.
We saw that the topological structure of
$QCD_{1+1}$ on a cylinder smears the infrared confinement mechanism for some of
the matter currents. We argued that if one attempts to promote this one
dimensional confinement mechanism to a three dimensional confinement
mechanism via the LF four Fermion instantaneous interaction,
as Wilson and Perry have
suggested, one must consider the topological effects.
The result in two dimension
was particular  to the cylinder topology of the space,
tempting one to argue that
this effect is irrelevant, but in
$QCD_{3+1}$ the topology is unavoidable.  Instantons exist and are
know to be essential in the solution of the $U(1)_A$ problem. Furthermore
the instantons that appear in LF
$QCD_{3+1}$ produce structure similar to those that smeared out the
confining instantaneous interaction in two dimensions.

\end{document}